\documentclass[journal=apchd5,manuscript=article,layout=twocolumn]{achemso}

\usepackage{graphicx}
\usepackage{dcolumn}
\usepackage{bm}

\usepackage[utf8]{inputenc}
\usepackage[T1]{fontenc}
\usepackage{mathptmx}
\usepackage{xcolor}
\usepackage{makecell}
\usepackage{amsmath}

\usepackage{cuted} 

\author{Mahtab A. Khan}
\email{mahtabahmad.khan@fuuast.edu.pk}
\affiliation{ 
NanoScience Technology Center and Department of Physics, University
of Central Florida, Orlando, FL 32826, USA
}%
\alsoaffiliation{Department of Applied Physics, Federal Urdu University of Arts, Science and Technology, Islamabad, Pakistan}%
\author{Jayden D. Craft}%
\affiliation{ 
NanoScience Technology Center and Department of Physics, University
of Central Florida, Orlando, FL 32826, USA
}%
\author{Hari P. Paudel}%
\affiliation{ 
National Energy Technology Laboratory, United States Department of Energy, 626 Cochran Mill Road, Pittsburgh, PA 15236, USA.
}%
\author{Yuhua Duan}%
\affiliation{ 
National Energy Technology Laboratory, United States Department of Energy, 626 Cochran Mill Road, Pittsburgh, PA 15236, USA.
}%
\author{Dirk R. Englund}%
\affiliation{ 
Department of Electrical Engineering and Computer Science, Massachusetts Institute of Technology, Cambridge, MA 02139, USA.
}%
\author{Michael N. Leuenberger}%
 \email{michael.leuenberger@ucf.edu}
\affiliation{ 
NanoScience Technology Center and Department of Physics, University
of Central Florida, Orlando, FL 32826, USA
}%
\alsoaffiliation{ 
College of Optics and Photonics (CREOL), University
of Central Florida, Orlando, FL 32826, USA.
}%

\date{\today}

\title{Er$_\mathrm{Al}$:Al$_2$O$_3$ for Telecom-Band Photonics: Electronic Structure and Optical Properties.}

\abbreviations{IR,NMR,UV}

\begin{document}





\begin{strip}
\maketitle
\begin{abstract}
Er-doped Al$_2$O$_3$ is a promising host for telecom-band integrated photonics. Here we combine \textit{ab initio} calculations with a symmetry-resolved analysis to elucidate 
substitutional Er on the Al site (Er$_\mathrm{Al}$) in $\alpha$-Al$_2$O$_3$. First-principles relaxations confirm the structural stability of Er$_\mathrm{Al}$. We then use the local trigonal crystal-field symmetry to classify the Er-derived impurity levels by irreducible representations and to derive polarization-resolved electric-dipole selection rules, explicitly identifying the symmetry-allowed $f$–$d$ hybridization channels. Kubo--Greenwood absorption spectra computed from Kohn--Sham states quantitatively corroborate these symmetry predictions. 
Furthermore, we connect the calculated intra-$4f$ line strengths to Judd--Ofelt theory, clarifying the role of $4f$–$5d$ admixture in enabling optical activity. Notably, we predict a characteristic absorption near 1.47~\textmu m (telecom band), relevant for on-chip amplification and emission. 
To our knowledge, a symmetry-resolved first-principles treatment of Er:Al$_2$O$_3$ with an explicit Judd--Ofelt interpretation has not been reported, providing a transferable framework for tailoring rare-earth dopants in wide-band-gap oxides for integrated photonics. Our results for the optical spectra are in good agreement with experimental data. 
\end{abstract}
\end{strip}

\section{Introduction}
Rare-earth (RE) doped materials have emerged as promising candidates for integrated photonics, providing an alternative to conventional III–V semiconductor compounds for on-chip light amplification and generation. These materials offer several notable advantages: they are highly compatible with silicon photonics platforms,\cite{Al2O3_Si_Interface,SMIT1986171} possess long excited-state lifetimes, and are easy to implement experimentally. RE doping has been realized in various semiconductors and insulators. Among the different rare-earth ions, erbium holds a significant advantage because it emits in the telecom wavelength range ($\sim$ 1.5$\mu$m), allowing signal transmission through optical fibers with minimal losses in modern fiber-optic technology.\cite{Bonneville:24,4810183}
\begin{figure*}[tbp]
	\begin{center}
		\includegraphics[width=7in]{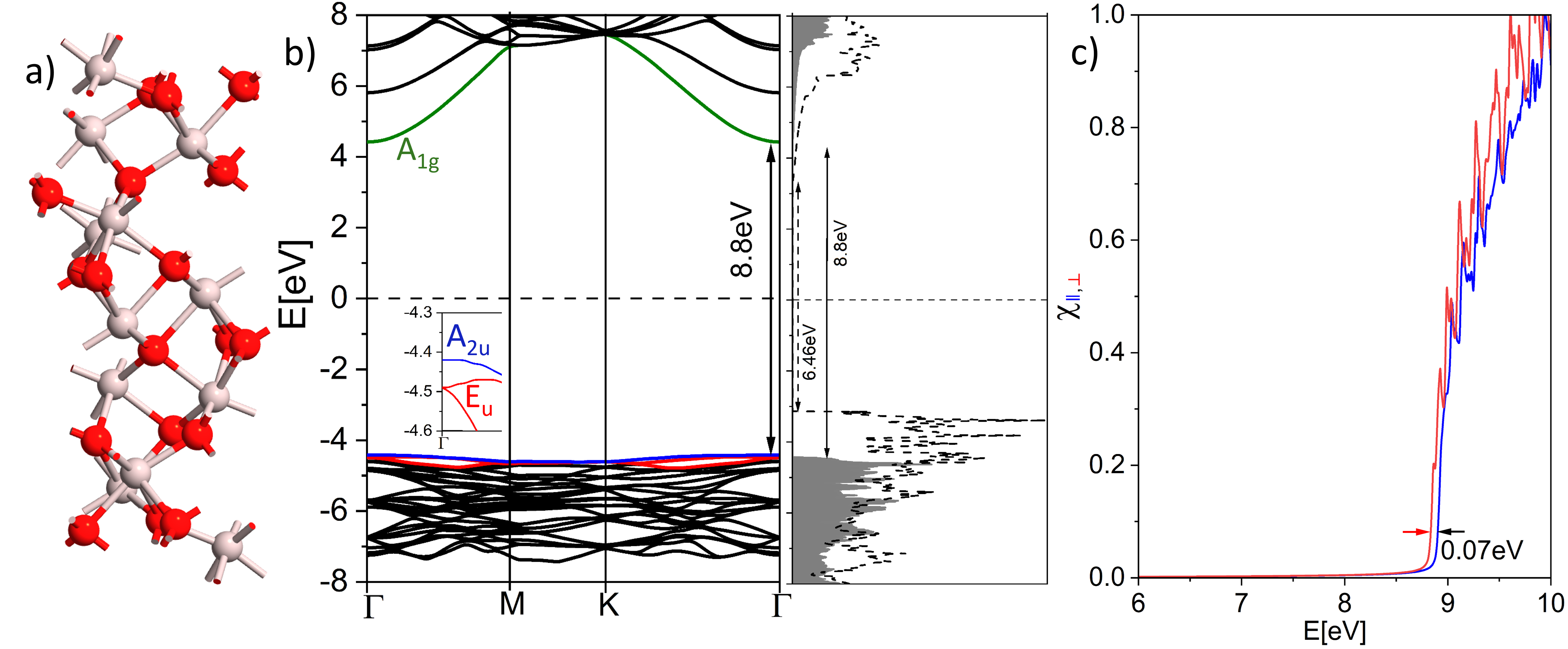}
	\end{center}
	\caption{
    a) Conventional cell of pristine $\alpha$-Al$_2$O$_3$ (space group R-3c) containing six formula units (30 atoms). Pink circles denote Al atoms; red circles denote O atoms.
    b) Electronic band structure of pristine $\alpha$-Al$_2$O$_3$, showing a direct gap at $\Gamma$ of E$_g=$ 8.8 eV. Symmetry labels at $\Gamma$ VB-1/-2 (degenerate) $\rightarrow$ E$_u$, VB $\rightarrow$ A$_{1u}$ and CB $\rightarrow$ A$_{1g}$. The corresponding DOS is shown: shaded-mGGA; dotted-GGA, which underestimates the gap (E$_g=$ 6.46 eV). The dashed line at 0 eV is the Fermi Energy E$_F$.  
    (c) $\alpha$-Al$_2$O$_3$ has two polarization-resolved onsets separated by 0.07 eV (z vs. x/y) as shown in the  susceptibility tensor $\chi_{\parallel,\perp}$ ($\chi_{\parallel}=\chi_{xx}+\chi_{yy}/2$ and $\chi_{\perp}=\chi_{zz}$).}
	\label{fig:Er_W_defect}
\end{figure*}

One of the main advantages of doping RE ions into various semiconductors and insulators is their unique property of having electrons in the unfilled 4f shell, which is strongly shielded from the crystal environment by the surrounding completely filled 5p and 5s shell. This property leads generally to high quantum yields, atom-like narrow bandwidths for optical transitions, long lifetimes, long decoherence times, high photostability, and large Stokes shifts. A prime example is Er-doped semiconductors and optical fibers that emit 1.5 $\mu$m light with ultra-narrow bandwidth and are therefore crucial for optoelectronic devices and optical telecommunication. 

To achieve maximum gain and improve the efficiency of Er-doped materials, it is essential to carefully consider factors such as the compatibility of Er impurities with the host material. SiO$_2$ is widely regarded as a reliable optical host material; however, it is not suitable for hosting Er impurities because of its inherently low solubility.\cite{KENYON2002225, Kik_Polman_1998} This limitation gives rise to undesirable transitions that significantly reduce optical gain while increasing noise in Er-doped systems. An equally critical challenge lies in the precise incorporation of impurities into the host: excessive doping concentrations promote non-radiative transitions, which in turn quench the absorption peaks and severely degrade device performance. Recently, it has been demonstrated that among various wide band gap materials, Al$_2$O$_3$ stands out as a suitable choice of host material for Er impurities due to its similarity with Er$_2$O$_3$ in terms of both valency and lattice constant.\cite{ronn2016atomic, Bradley:10, Al2O3_Si_Interface, SMIT1986171}

Er-doped Al$_2$O$_3$ (Er$\rm _{Al}$:Al$_2$O$_3$) underpins on-chip light sources and amplifiers in the telecom band, yet a symmetry-resolved, first-principles framework for its optical activity is lacking. In particular, the Er impurity levels in the local $C_{3v}$ crystal field have not been systematically classified by irreducible representations, the ensuing electric-dipole selection rules have not been established from group theory, and the connection to Judd--Ofelt\cite{Judd, Ofelt} (JO) theory has not been articulated within an \textit{ab initio} context. Here we address this gap by combining density-functional theory with a comprehensive group-theoretical analysis to identify the $A_{1}$/$E$ manifolds, derive polarization-resolved dipole selection rules, and quantify the role of symmetry-allowed $4f$--$5d$ admixture. We validate these predictions against Kubo--Greenwood absorption spectra computed from Kohn--Sham states and show how the JO framework naturally emerges from the calculated line strengths. This symmetry-grounded approach links crystal-field physics to device-relevant spectral features in Er:Al$_2$O$_3$.

 In free ions, $4f\!\to\!4f$ electric–dipole transitions are parity-forbidden; in crystals they acquire intensity via opposite-parity admixture, consistent with the Judd–Ofelt mechanism. 
For Er substituting Al in $\alpha$-Al$_2$O$_3$, local relaxation lowers the site symmetry to noncentrosymmetric $C_{3v}$, allowing odd crystal-field components that mix $4f$ and $5d$ character within matching irreducible representations ($A_1$–$A_1$, $E$–$E$). 
This $4f$--$5d$ admixture activates nominally forbidden $4f$--$4f$ lines and renders them optically active.
\hfill

Building on methodology we previously validated for impurity/defect-bound optical transitions,\cite{ER_W_khan,khan2022first,Khan2017,Erementchouk2015} we compute the symmetry-resolved electronic structure and polarization-dependent absorption of substitutional Er (Er$_\mathrm{Al}$) in $\alpha$-Al$_2$O$_3$ from first principles. Our analysis predicts a telecom band resonance near 1.47~\textmu m and attributes it to transitions within the $A_{1}$/$E$ manifolds, activated by symmetry-allowed $4f$--$5d$ admixture. While the wavelength agrees with prior spectroscopy on Er-doped glasses and Er:Al$_2$O$_3$,\cite{Er_Glass_Laser,Malinowski2000,ronn2016atomic,aisaka2008enhancement,xiao2022design} the present work provides, to our knowledge, the first \textit{ab initio}, symmetry-resolved assignment of its microscopic origin and polarization selection rules in this host.
 
 In this work, by combining \textit{ab initio} Density Function Theory (DFT), Kubo--Greenwood susceptibility formula, and a $C_{3v}$ group-theoretical analysis, we demonstrate:
(i) the presence of Er-derived localized states within the bandgap of $\alpha$-Al$_2$O$_3$;
(ii) narrow, atom-like intra-$4f$ optical transitions; 
(iii) polarization-resolved electric-dipole selection rules dictated by the local symmetry; and
(iv) a Judd--Ofelt (JO) interpretation whereby electric–dipole activity of intra-$4f$ lines arises from symmetry-allowed $4f$--$5d$ admixture.

\begin{figure*}[hbt]
	\begin{center}
		\includegraphics[width=7in]{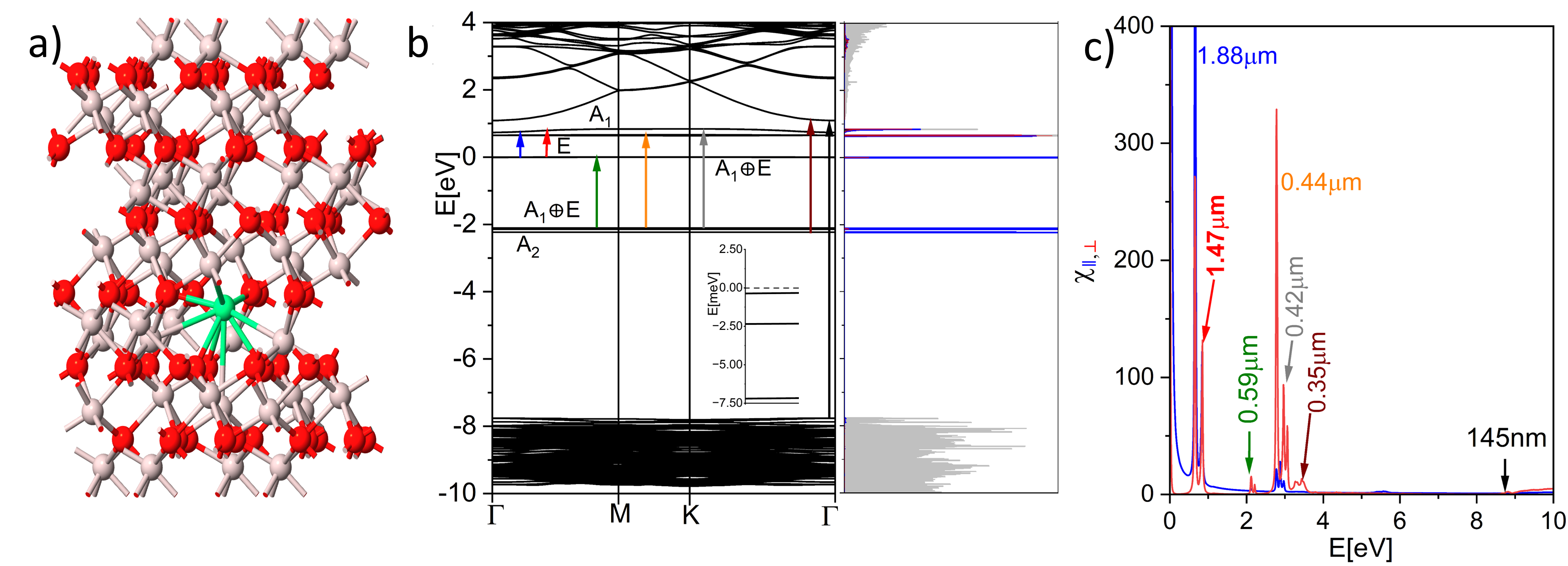}
	\end{center}
	\caption{a) Structure of Er$\rm _{Al}$ impurity in 2$\times$2$\times$1 Al$_2$O$_3$. The minimum impurity seperation is $\sim$ 10 \AA~in this configuration. b) Band structure of 2$\times$2$\times$1 of Er$\rm _{Al}$:Al$_2$O$_3$. Vertical arrows indicate optical transitions corresponding to resonances in the electric susceptibility plot as shown in c).Impurity states in the band structure belonging to the A$_{1}$, A$_2$ and E (doubly degenerate) representations of the C$_{3v}$ point group. The inset magnifies the region near the Fermi level, at $\Gamma$-point; the dashed line at 0~eV marks $E_F$. Density of states plot is also shown in (b), shaded grey region shows the total density of states and the blue (red) curve shows the projected density of states of $f$- ($d$-) orbital of the Er atom, of $2\times2\times1$ supercell of Er$_{\rm Al}$:Al$_2$O$_3$. The dispersionless localized impurity states (LIS) corresponding to the $f$-orbitals of Er are clearly visible as dispersionless (localized) states, lying within the bandgap region of Al$_2$O$_3$. Vertical arrows show some of the dipole allowed transitions. Allowed transitions appears as resonance peaks in the susceptibility tensor $\chi_{\parallel,\perp}$ shown in (c).}
	\label{fig:bandstructure_Ho_W}
\end{figure*}
 \section{Band structure}
 \label{sec:bandstructure}
 
 \subsection{Numerical Analysis}
  All numerical calculations are carried out using DFT. The generalized gradient approximation (GGA) is used  with the Perdew-Burke-Ernzerhof (PBE) parametrization\cite{PhysRevLett.77.3865} of the correlation energy for geometrical optimization. The calculations are implemented within the Synopsis Atomistix Toolkit (ATK) 2021.06.\cite{QW_1} 
 The periodic structure of the supercell allows one to characterize the electron states by the band structure $\epsilon_n(\mathbf{k})$, where $\mathbf{k}$ is the vector in the first Brillouin zone of the supercell and $n$ enumerates different bands. Semilocal GGA functionals are known to underestimate band gaps of semiconductors and insulators, largely due to self-interaction errors and the missing derivative discontinuity of the exchange–correlation functional. Quasiparticle $GW$ calculations can correct this deficiency, but they are computationally prohibitive for the large supercells (hundreds of atoms) considered here. Hybrid functionals—mixing a fraction of exact (Hartree–Fock) exchange with a semilocal functional—often improve gap predictions, but at a substantially higher computational cost. As a practical and accurate alternative, meta-GGA functionals\cite{meta_GGA} (which depend on the density, its gradient, and the kinetic-energy density) frequently yield better band gaps and more realistic placement of localized states than GGA, at a cost far lower than hybrids. In this work we therefore employ a meta-GGA functional to compute the band structure of $\alpha$-Al$_2$O$_3$, obtaining a band gap close to experiment.
 
 We first calculated the band structure of pristine Al$_2$O$_3$. The formula unit of Al$_2$O$_3$ contains five atoms. \textcolor{black}{$\alpha$-Al$_2$O$_3$ crystallizes in the corundum structure (space group $R\bar{3}c$, No.~167).  The primitive rhombohedral unit cell contains one lattice point with a basis of two formula units  (Al$_4$O$_6$), i.e., 10 atoms. Equivalently, the structure may be described in the conventional hexagonal setting, which comprises six close-packed oxygen layers stacked along $c$; the smaller  Al cations occupy two-thirds of the octahedral interstitial sites. In this representation the conventional  hexagonal cell\cite{galasso2013structure} has three times the volume of the primitive rhombohedral cell and therefore contains  six formula units (30 atoms), as shown in Fig.~\ref{fig:Er_W_defect}(a). Pristine $\alpha$-Al$_2$O$_3$ belongs to the $D_{3d}$ point group and is centrosymmetric.} 
 With inversion and time-reversal symmetry, the host bands of $\alpha$-Al$_2$O$_3$ remain spin-degenerate and spin–orbit coupling (SOC) effects on the Al/O edges are small relative to the wide gap. By contrast, SOC on Er $4f$ states sets the $J$-multiplet fine structure. Our $C_{3v}$ analysis targets the symmetry labels and polarization selection rules that govern allowed electric–dipole transitions; SOC refines the fine structure without altering these rules.
Therefore, we performed unpolarized DFT calculations using a meta-GGA hybrid exchange correlation functional to obtain the band structure, density of states, and electric susceptibility. We obtain a band gap (direct) of 8.8 eV at the $\Gamma$ point, which is in very close agreement with experimental values. \cite{Exp_BG_1, Exp_BG_2} At the $\Gamma$-point Al$_2$O$_3$ has two polarization-resolved onsets separated by 0.07 eV (z vs. x/y). At the $\Gamma$-point the valence-band maximum transforms as A$_{2u}$ and the conduction-band minimum as A$_{1g}$ IR's (irreducible representations of D$_{3d}$). Consequently, the direct A$_{2u}$ $\rightarrow$ A$_{1g}$ transition is electric-dipole allowed for z-polarization ($\pi$), since A$_{2u}$ $\otimes$ A$_{2u}$ $\otimes$ A$_{1g}$ $=$ A$_{1g}$. The next two valence bands (VB–1 and VB–2) form a degenerate E$_u$ doublet at $\Gamma$; there transitions to A$_{1g}$ conduction band are allowed for in-plane (x,y) polarization ($\sigma$), because E$_{u}$ $\otimes$ E$_{u}$ $\otimes$ A$_{1g}$ $=$ A$_{1g}$ $\oplus$ A$_{2g}$ $\oplus$ E$_{g}$. \textcolor{black}{The matrix elements of the susceptibility tensor in three dimensions can be evaluated using the Kubo–Greenwood formula for the electric susceptibility 
\begin{equation}\label{eq:KGW}
\chi_{ij}(\omega)=\frac{e^{2}}{\hbar m_{e}^{2}V}\sum_{uv\bf{k}}\frac{f_{u\bf{k}}-f_{v\bf{k}}}{\omega_{uv}^2({\bf{k}})[\omega_{uv}({\bf{k}})- \omega-i\Gamma/{\hbar}]}p_{uv}^{i}p_{vu}^{j},
\end{equation}  
where $p_{pq}^{j}=\langle u{\bf{k}}|p^{j}|v\bf{k}\rangle$ is the $j$th component of dipole matrix element between states $u$ and $v$ with the Bloch state $\langle\bf{r}$$|u$$\bf{k}\rangle= \psi_{{u}\mathbf{k}}(\bf{r})$, $V$ the volume of the crystal, $f$ the Fermi function, and $\Gamma$ the broadening, which is set to be 0.01 eV. When recasting the $\mathbf{k}$-point sum in Eq.~(\ref{eq:KGW}) as an energy integral, the relevant quantity is the joint density of states (JDOS), which counts pairs of initial and final states at the same $\mathbf{k}$ whose energy difference matches $\hbar\omega$, and it must be weighted by the dipole matrix elements. In this representation, the imaginary part of the response can be written as $\mathrm{Im}\,\chi_{ij}(\omega) = \frac{\pi e^{2}}{\hbar m_{e}^{2}}\sum_{uv}\omega^{-2}\, \mathcal{J}^{ij}_{uv}(\omega)$, where $\mathcal{J}^{ij}_{uv}(\omega)$ is the matrix-element-weighted JDOS. The optical response is relatively weak (Fig.~\ref{fig:Er_W_defect} (c)), as the density of states (DOS) near the conduction-band edge is small.}

\begin{table*}[h]%
\centering%
\begin{tabular}{ |c |c |c |c |c |c |c |c |}\hline
$D_{3d}$    &    $E$   &   $2C_{3}$   &   $3C^{\prime}_{2}$  &  $i$  &  $2S_{6}$   &   $3\sigma_{d}$  &  Linear Rotations\\ 
\hline
$A_{1g}$    & 1   & 1    & 1    & 1    &    1    &  1      &                  \\
 $A_{2g}$   & 1   & 1    & $-$1  &  1  &    1    & $-$1    &       R$_z$             \\
 $E_{g}$    & 2   & $-$1  & 0    &  2     &      $-$1       & 0   &  (R$_x$, R$_y$)                     \\
  $A_{1u}$  & 1        & 1        & 1      & $-$1     &      $-$1        &  $-$1     &                    \\
 $A_{2u}$                    & 1        & 1                      & $-$1                  & $-$1                            &         $-$1                          &  1        & p$_z$                \\
 $E_{u}$                   & 2        & $-$1                      & 0                  & $-$2                            &         1                          &  0              &  (p$_x$, p$_y$)               \\
\hline\end{tabular}
\caption{}
\label{table_D_3h}
\begin{tabular}{ |c |c |c |c |c |c |c |}\hline
$D_{3d}$    &    $A_{1g}$   &   $A_{2g}$   &   $E_{g}$  &  $A_{1u}$  &  $A_{2u}$   &   $E_{u}$  \\ 
\hline
$A_{1g}$    &   &    &     &     &    $\pi$    &  $\sigma$           \\
\hline
 $A_{2g}$   &    &     &   &  $\pi$  &        & $\sigma$                 \\
 \hline
 $E_{g}$    &    &   &     &   $\sigma$    &    $\sigma$        & $\sigma, \pi$                \\
 \hline
  $A_{1u}$  &         & $\pi$        & $\sigma$      &      &             &                           \\
  \hline
 $A_{2u}$    & $\pi$        &        & $\sigma$      &      &         &                         \\
 \hline
 $E_{u}$   & $\sigma$    & $\sigma$    & $\sigma,\pi$   &   &           &                         \\
\hline\end{tabular}
\caption{Character table of the point symmetry group $D_{3d}$ and Dipole allowed transitions of symmetry group $D_{3d}$. $\sigma$ denotes the allowed dipole transitions along the basal plane xy while $\pi$ denotes transitions along z-axis}
\label{D_3h_DSR}
\end{table*}

 For an Er$\rm _{Al}$ impurity, we consider a $2 \times 2 \times 1$ (see Fig.~\ref{fig:Er_W_defect}) supercell having 120 atoms with a doping percentage of 0.83$\%$. \textcolor{black}{In pristine $\alpha$-Al$_2$O$_3$ (space group $R\bar{3}c$), the two Al atoms in the primitive rhombohedral  cell belong to a single crystallographic Al site (same Wyckoff orbit), so all Al sites are symmetry-equivalent.} The Brillouin zone of the supercell is sampled by a 4$\times$4$\times$2 $k$-mesh. All the structures are geometrically optimized with a force tolerance of $0.01$ eV/\AA. First we calculate the stability of the Er impurity in Al$_2$O$_3$. The formation energy $E_{f,\text{imp}}$ of Er$_{\text{Al}}$:Al$_2$O$_3$ is defined with respect to the chemical potentials of Al and Er, and is obtained as the energy difference between the total energy of the Er$_{\text{Al}}$:Al$_2$O$_3$ structure and the corresponding combination of reference reservoirs for Er and Al. The formation energy can be expressed as
 \begin{align*}
E_{f,\mathrm{imp}}\left(\mathrm{ErAl_{47}O_{72}}\right)
  &= E_{\mathrm{tot}}\!\left(\mathrm{ErAl_{47}O_{72}}\right)
     - E_{\mathrm{tot}}\!\left(\mathrm{Al_{48}O_{72}}\right) \\[2pt]
  &\quad - \mu_{\mathrm{Er}} + \mu_{\mathrm{Al}} \, .
\end{align*}
 where, $\rm E_{tot}(\rm Er_{Al}:(Al_2O_3))$ is the total energy of the $\rm Er_{Al}:Al_2O_3$ structure, $\mu_{X}$ (X = Er, Al) are the chemical potentials of atom X, calculated from their bulk counter parts. The calculated value of E$\rm _{f,imp}$ per defect for  Er$\rm _{Al}:(Al_2O_3)$ is 2.638 eV. The calculated formation energy corresponds to the substitution of one Al atom in Al$_2$O$_3$ by an Er atom, referenced to the chemical potentials of elemental Al and Er. \textcolor{black}{An alternative way is to calculate the formation energy of Er$_{Al}$:Al$_2$O$_3$ with respect to the gas phase\cite{Impurity_Enginneered_Graphullerene} of Er, Al and O atoms by using the following relation
 \begin{align*}
E_f^{\text{gas}}\!\left(\mathrm{ErAl_{47}O_{72}}\right)
  &= E_{\text{tot}}\!\left(\mathrm{ErAl_{47}O_{72}}\right) - \mu^{\text{gas}}_{\mathrm{Er}} \\
  &\quad - 47\,\mu^{\text{gas}}_{\mathrm{Al}} - 72\,\mu^{\text{gas}}_{\mathrm{O}} \, .
\end{align*}
where $\mu^{gas}\rm _{X}$ (X$=$Er, Al, O) is the chemical potential of X atom in the gas phase. This formulation enables a consistent comparison of stability across pristine and Er doped Al$_2$O$_3$ by anchoring all formation energies to a common gas-phase reference. The calculated value $E^{gas}_f(\rm ErAl_{47}O_{72})=-$14.83 eV per atom indicates that the Er$\rm _{Al}$:Al$_2$O$_3$ structure is thermodynamically stable. It should be noted that Er-doped Al$_2$O$_3$ samples have been realized experimentally.\cite{4810183, Vazquez-Cordova:14}}

In Er-doped Al$_2$O$_3$, Er$^{3+}$ is an isovalent substitute for Al$^{3+}$, so no charge compensation is required. Despite its larger ionic radius (Er$^{3+}$ $\approx$ 0.89 \AA, CN$=$6) compared with Al$^{3+}$ ($\sim$ 0.535 \AA), the Al$_2$O$_3$ lattice can accommodate Er$^{3+}$ on the octahedral cation site (local C$_{3v}$ symmetry) at dilute concentrations, with only local distortions. The global crystal symmetry (R$-$3c) is preserved, while the non-centrosymmetric local site enables weakly allowed 4f$-$4f transitions via f$-$d mixing. At higher Er loadings, strain and Er–Er interactions promote photoluminescence quenching and clustering toward Er$_2$O$_3$, so device design typically targets sub-percent atomic fractions.\cite{ronn2016atomic}
 
 Fig.~\ref{fig:bandstructure_Ho_W} shows the band structure of Er$\rm _{Al}$:Al$_2$O$_3$. Al$_2$O$_3$ has a wide band gap and the presence of Er$\rm _{Al}$ impurity in Al$_2$O$_3$ leads to localized impurity states (LIS) which can be seen as dispersionless electronic states within the band gap region of Al$_2$O$_3$. Vertical arrows show some of the allowed optical transitions between different $f-$orbitals of Er$\rm _{Al}$ impurity, observed in the optical spectra (see Fig.~\ref{fig:bandstructure_Ho_W} c).
 
\begin{table*}[hbt]%
\centering%
\begin{tabular}{ |c |c |c |c |c |c |c |}\hline
C$\rm _{3}$    &    $E$   &   $C_{3}$   &   3$\sigma_v$   &  \makecell{Linear\\Rotations}  & \makecell{quadratic\\
functions} & \makecell{Cubic\\ functions}\\ 
\hline
A$\rm _1$    & 1   & 1    & 1         &  z  & \makecell{z$^2$\\x$^2$$+$y$^2$} & z$^3$, x(x$^2$-3y$^2$), z(x$^2$+y$^2$)           \\
\hline
A$\rm _2$    & 1   & 1  & -1   &  R$\rm _{z}$ &     -  &    y(3x$^2$-y$^2$)       \\
\hline
E    & 2   & 1  & 0   &  \makecell{(x, y)\\(R$\rm _x$, R$\rm _y$)}  &  \makecell{(x$^2$-y$^2$, xy)\\ (xz, yz)}     &  \makecell{(xz$^2$, yz$^2$) $\lbrack$ xyz, z(x$^2$-y$^2$)$\rbrack$\\$\lbrack$x(x$^2$+y$^2$), y(x$^2$+y$^2$)$\rbrack$} \\
\hline\end{tabular}
\caption{ }
\label{table_C_3}
\centering%
\begin{tabular}{ |c |c |c |c |}\hline
$C_{3}$    &    A$_1$   &   A$_2$   &   E   \\ 
\hline
A$_1$    & $\pi$   &     & $\sigma$                        \\
\hline
A$_2$    &   & $\pi$  & $\sigma$          \\
\hline
E  & $\sigma$  & $\sigma$   & $\pi,\sigma$\\
\hline\end{tabular}
\caption{Character table of the group $C\rm _{3v}$ and Dipole selection rules for C$\rm _{3v}$ symmetry.}
\label{table_C_3_DSR}
\end{table*}
 When considering impurities in a crystal, the LIS transform according to the IRs of the symmetry group of the crystal site in which the impurity resides. While the translational symmetry of the crystal is broken, point group symmetries are partially or completely preserved. The Er$\rm _{Al}$ impurity locally breaks the D$\rm _{3d}$ symmetry of Al$_2$O$_3$ and is therefore represented locally by a lower symmetry group C$\rm _{3v}$ the irreducible representations (IRs) of which are shown in Table~\ref{table_C_3}. We can identify the IR's of the impurity states by plotting the Bloch states at the $\Gamma$-point. 

\begin{figure}
\centering
\includegraphics[width=2.3in]{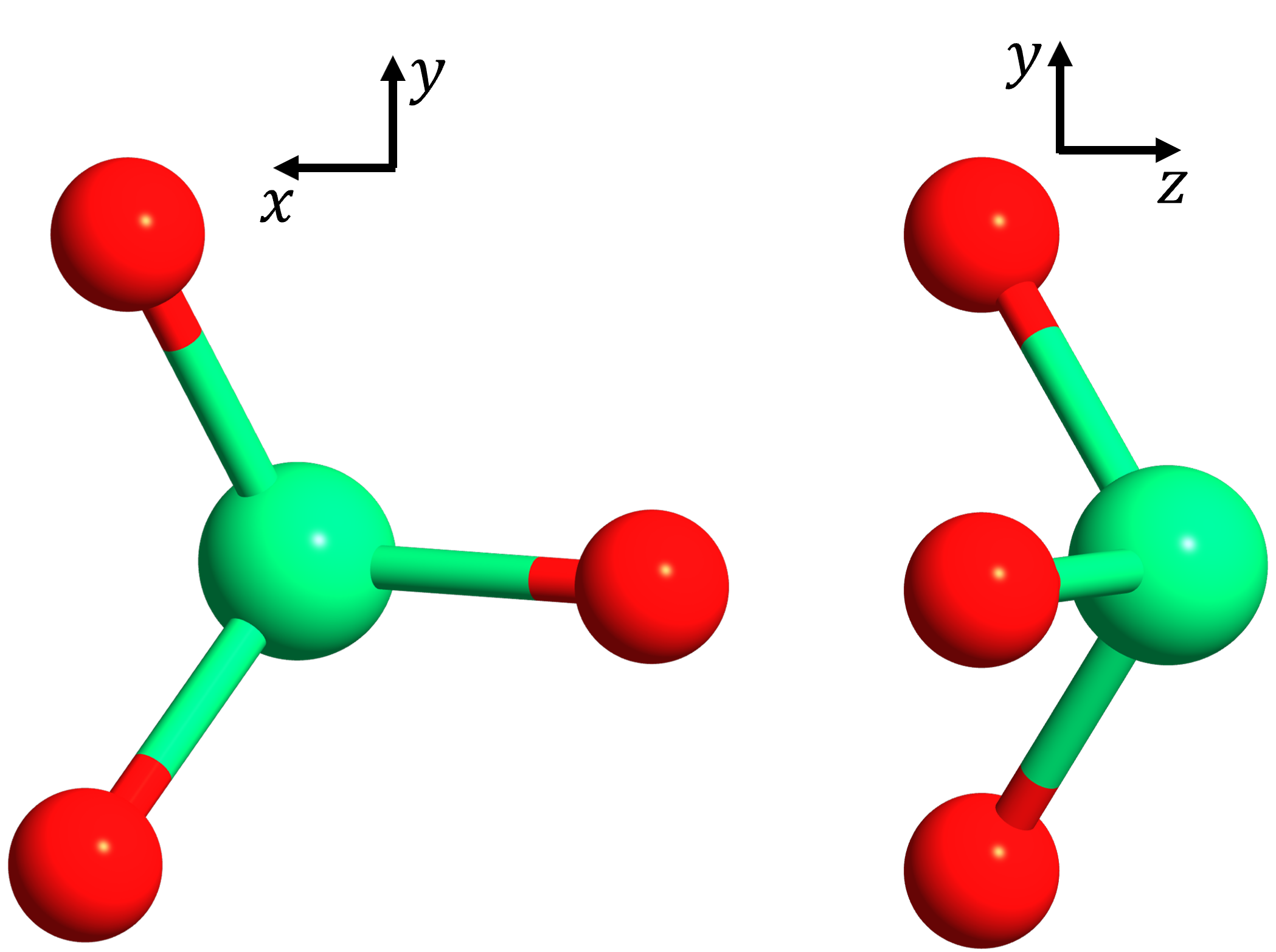}
\caption{Er$\rm _{Al}$ impurity with nearest neighbor O atoms, forming C$\rm _{3v}$ symmetry.}
\label{fig:C3v_Symmetry}
\end{figure}

 \paragraph{Er$_{\mathrm{Al}}$ in $\alpha$-Al$_2$O$_3$: trigonal crystal field and $f$-level splitting.}
Er$^{3+}$ substituting Al$^{3+}$ in Al$_2$O$_3$ occupies the octahedral cation site in a trigonal environment. In a local frame with $\hat z \parallel \hat C_{3}$, the impurity is described by
\begin{align}
H &= H_{\text{host}} + H_{\text{CF}}(\mathrm{Er}), \\[6pt]
H_{\text{CF}} &= \sum_{k=2,4,6} \;\sum_{q=0,\pm 3,\pm 6} B_k^{q}\, C_k^{q}.
\end{align}
where $C_k^{q}$ are spherical-tensor (Wybourne/Racah) operators acting within the $l=3$ $f$-manifold and $B_k^{q}$ are crystal-field parameters allowed by the C$\rm _{3v}$ symmetry (only $q=0,\pm3,\pm6$ survive the threefold rotation). Reducing the seven-dimensional $f$-orbital representation to the effective trigonal site symmetry of the Al$_2$O$_3$ cage (conveniently taken as $C_{3v}$ for selection rules) gives
\[
\Gamma_{f}\!\downarrow C_{3v} \;=\; 2A_{1} \,\oplus\, A_{2} \,\oplus\, 2E 
\;=\; 3A \,\oplus\, 2E,
\]
hence the seven $f$-orbitals split into three singlets and two doubly degenerate $E$ levels at an Er$_{\mathrm{Al}}$ site with local trigonal symmetry. It should be noted that the above description is true only when $f$-orbitals are involved. However it can be seen in FIG.~\ref{fig:bandstructure_Ho_W} (b) that substantial admixture of 5$d$-orbitals is also present in $f$-orbitals. In $C_{3v}$ symmetry, the $f$- and $d$-orbital manifolds decompose as
\begin{align}
f:&\; 2A_{1}\oplus A_{2}\oplus 2E \quad (7~\text{orbitals}), \\[6pt]
d:&\; A_{1}\oplus 2E \quad (5~\text{orbitals}).
\end{align}

Allowing symmetry-preserving $f$--$d$ hybridization (only between states
of the same irrep: $A_{1}\!\leftrightarrow\!A_{1}$, $E\!\leftrightarrow\!E$), the combined
orbital content is
\[
(f\oplus d)\big|_{C_{3v}} = 3A_{1}\oplus A_{2}\oplus 4E,
\]
which corresponds to a total of
\[
3\times 1 \;+\; 1\times 1 \;+\; 4\times 2 \;=\; 12
\]
orbital states. The $A_{2}(f)$ level has no $d$-orbital
counterpart in $C_{3v}$ and therefore remains essentially unmixed to first order. Out of these 12 states most lie in-gap; two reside well inside the conduction band.

  \begin{figure*}
\centering
\includegraphics[width=7.1in]{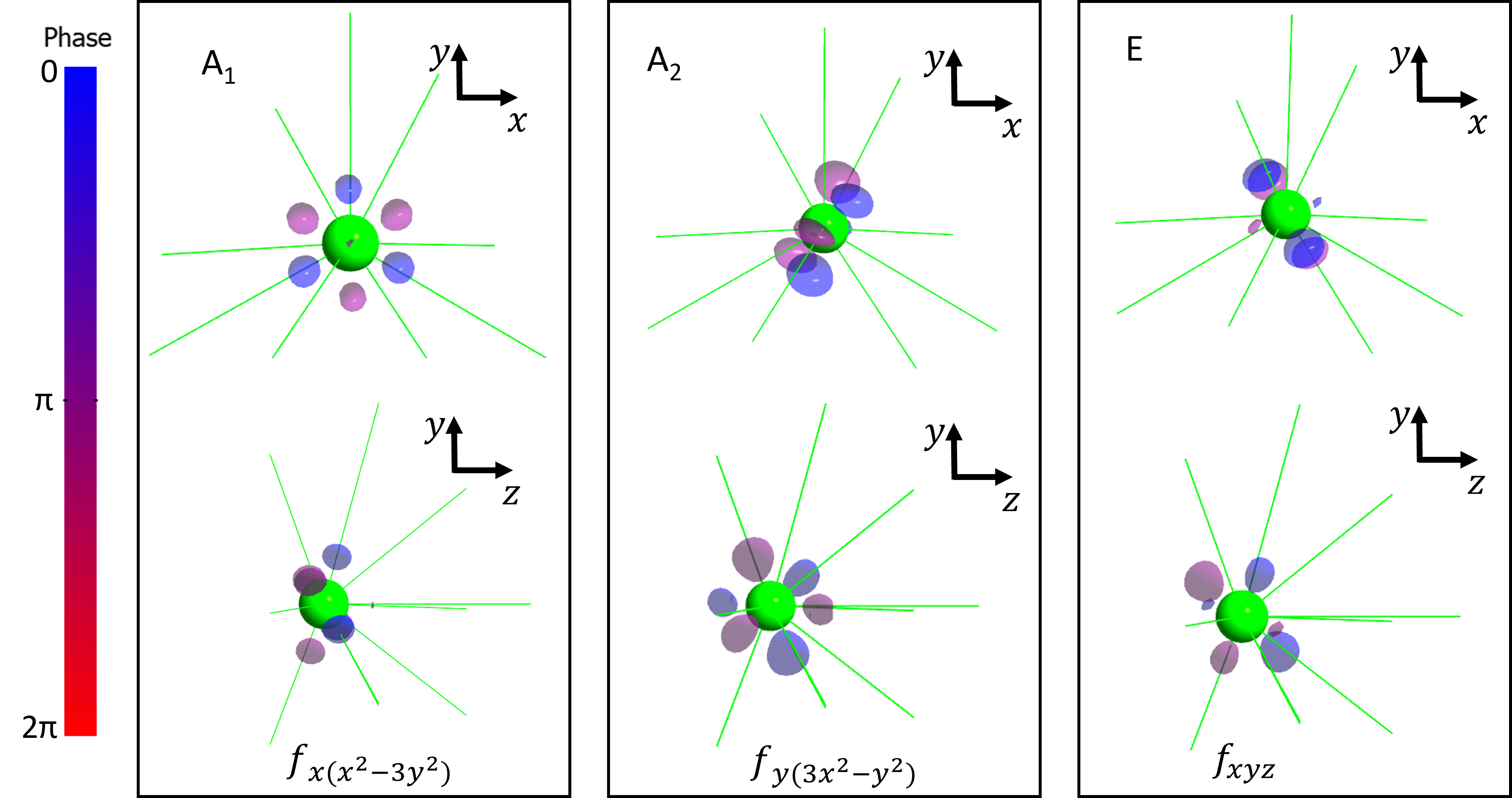}
\caption{Examples of the Bloch states at $\Gamma$-point for the Er$\rm _{Al}$ impurity in 2$\times$2$\times$1 super cell of Al$_2$O$_3$. The green ball shows the Er$\rm _{Al}$ impurity atom.}
\label{fig:Bloch_states}
\end{figure*}

\section{Optical Response}
\label{sec:optical}
The presence of localized in-gap states (LIS) in the band structure gives rise to sharp peaks in the susceptibility tensor, which we calculate using the Kubo--Greenwood (KG) formula, Eq.~(\ref{eq:KGW}). In Fig.~\ref{fig:bandstructure_Ho_W} (c) results for the $\chi_{\parallel,\perp}$ component of the electric susceptibility tensor, which corresponds to the z-polarization, is presented for Er$\rm _{Al}$:Al$_2$O$_3$. The electric susceptibility provides valuable insights into the optical selection rules for transitions between states across the Fermi level. We are interested in transitions involving states with energy near the gap edges or inside the gap. Appearance of impurity states inside the band gap $E_{g}$ or close to the band edges leads to the resonances at single frequency $\hbar\omega_{uv}=|\varepsilon_{u}-\varepsilon_{v}|$, where $\varepsilon_u$ is the eigenenergy of the Bloch state $\psi_{u\bf{k}}(\bf{r})$. The dipole matrix element $p_{uv}^{j}$ determines the strength of an optical transition and whether it is allowed or prohibited by symmetries.

Radiative transitions between $f$-orbitals have been the subject of extensive research since the 1960s, as they are nominally dipole-forbidden due to electric dipole selection rules. The Judd–Ofelt theory\cite{Judd, Ofelt} accounts for these transitions by incorporating odd-parity terms in the crystal field Hamiltonian, which can exist only in non-centrosymmetric crystal environments. Using second-order perturbation theory, it has been shown that $f$-orbitals can mix with 5$d$-orbitals, i.e., $\langle f | \hat{H}_{\text{cry}} | d \rangle \neq 0$, only when $\hat{H}_{\text{cry}}$ contains odd-parity components, thereby enabling radiative transitions. In the present case, the Er impurity in Al$_2$O$_3$ resides at a Al with C$_{3v}$ symmetry, which lacks inversion symmetry, thereby permitting odd terms in the crystal field Hamiltonian. The crystal field potential is given by $\hat{H}_{\mathrm{CF}} = B_2^0 , \hat{O}_2^0 + B_3^3 , \hat{O}_3^3 + B_4^0 , \hat{O}_4^0 + \dots$. The odd-parity term $B^3_3$ , $\hat{O}^3_3$ arises solely due to the non-centrosymmetric nature of the C$_{3}$ point group and is responsible for enabling coupling between the 4$f$- and 5$d$- orbitals of Er, as evidenced by the blue and red curves in the density of states as shown in Fig.~\ref{fig:bandstructure_Ho_W} (b). The largest $4f$--$5d$ admixture occurs in the electronic states at $0.66$\,eV and $0.80$\,eV; a moderate component is present at the Fermi level ($E_F$), and a smaller but finite contribution appears in the states at $-2.12$\,eV.
 On the basis of Judd--Ofelt analysis, stronger $4f$--$5d$ admixture yields relatively more pronounced peaks in $\chi(\omega)$. Consistent with this expectation, the resonances observed at $1.88\,\mu\mathrm{m}$, $1.47\,\mu\mathrm{m}$, $0.44\,\mu\mathrm{m}$, and $0.42\,\mu\mathrm{m}$—assigned to transitions with substantial $4f$--$5d$ admixture—are markedly more pronounced in the susceptibility tensor.

The appearance of LIS inside the band gap leads to sharp resonances in electric susceptibility $\chi$  at frequencies corresponding to the energy differences between LIS. In particular we are able to identify a signature resonance peak of Er at 1.47$\mu$m, which is in very good agreement with the absorption peak found in Er glass lasers\cite{Er_Glass_Laser} and also in recently realized Er-doped single layer MoS$_2$.\cite{YANG2007207} Group theory provides an excellent framework to investigate electronic 
transitions both within the dipole approximation and beyond. Within the  dipole approximation, a transition is considered allowed when the matrix element $p_{uv}^{j}$ transforms according to the totally symmetric irreducible representation of the symmetry group of the supercell. 
In this context, the initial state $|v\mathbf{k}\rangle$, the final state $|u\mathbf{k}\rangle$, and the momentum operator $p^{j}$ transform according to the irreducible representations $\Gamma(|v\mathbf{k}\rangle)$, $\Gamma(|u\mathbf{k}\rangle)$, and $\Gamma(p^{j})$, respectively. The electric dipole transition between two states is allowed if the direct product 
\[
\Gamma(|v\mathbf{k}\rangle) \otimes \Gamma(p^{j}) \otimes \Gamma(|u\mathbf{k}\rangle)
\]
contains the identity representation $\Gamma(I)$ in its decomposition. 
For example, in the C$\rm _{3v}$ point group, the identity representation 
corresponds to A$_{1}$.
 This is strictly related to the polarization of the radiation. One needs to consider separately the in plane and out of plane components of $p_{vu}^{j}$ because they transform according to different IRs of C$_{3v}$. The selection rules for electric dipole transitions for IRs are summarized in Table~\ref{table_C_3_DSR}. 

We want to comment about the device relevance.
The prominent 1.47\,\textmu m resonance observed in both the out-of-plane ($\chi_{\perp}$) and in-plane ($\chi_{\parallel}$) responses indicates mixed $A_{1}$/$E$ character at the Er$_\mathrm{Al}$ site. In planar Al$_2$O$_3$ photonics, this enables polarization-diverse coupling: $A_{1}$-like components couple efficiently to TM-like modes (large $E_z$), while $E$-like components couple efficiently to TE-like modes (in-plane
$\mathbf{E}$). Consequently, the same spectroscopic feature can be harnessed in either TE or TM device families by tailoring field profiles (waveguide height/aspect ratio, slot geometries) or by orienting microresonators (rings/disks) to favor the desired polarization. The mixed character also suggests practical routes to polarization control: cavity designs that enhance $E_z$ selectively boost the $A_{1}$ contribution, whereas high-confinement TE modes emphasize the $E$ contribution, enabling dichroic operation and polarization-multiplexed amplification/emission near 1.47\,\textmu m. Since the degree of optical activation scales with symmetry-allowed $4f$--$5d$ admixture, modest engineering of the local crystal field (strain, co-dopants, composition) provides a symmetry-guided knob to optimize coupling without sacrificing spectral selectivity.

\section*{Conclusions}
We have presented a symmetry-resolved, first-principles framework for Er-doped $\alpha$-Al$_2$O$_3$ that connects local crystal-field physics to device-relevant optical response. 
Substitutional Er$_\mathrm{Al}$ is structurally stable and yields Er-derived localized states within the wide bandgap. 
Using the $C_{3v}$ site symmetry, we classified the impurity levels into $A_{1}$ and $E$ manifolds and derived the corresponding polarization-resolved electric–dipole selection rules. 
Kubo--Greenwood spectra computed from Kohn--Sham states corroborate these symmetry assignments and reveal a prominent telecom band feature near 1.47~\textmu m. 
Consistent with the Judd--Ofelt mechanism, line strengths increase with symmetry-allowed $4f$--$5d$ admixture ($A_{1}\!\leftrightarrow\!A_{1}$, $E\!\leftrightarrow\!E$), while the $A_{2}$ level remains essentially dark. 
Together, these results provide actionable design rules for Er:Al$_2$O$_3$: target IRs that maximize opposite-parity mixing, align device polarization with $A_{1}/E$ characters, and engineer the local environment to tune $f$–$d$ hybridization.
This framework is transferable to other rare-earth dopants and wide-band-gap oxides, offering a microscopic route to symmetry-guided photonic functionality.

\begin{acknowledgement}
J. D. C. acknowledges support by an appointment with the AMMTO Summer Internships program sponsored by the U.S.
Department of Energy (DOE), EERE Advanced Materials and Manufacturing Technologies Office (AMMTO). J. D. C. also acknowledges support by the DOE fellowship Office of Science Graduate Student Research (SCGSR) Program. 
M. N. L. and D. R. E. acknowledge support by the Air Force Office of Scientific Research (AFOSR) under award no. FA9550-23-1-0472.
\end{acknowledgement}


\bibliographystyle{apsrev4-1}
\bibliography{bibliography.bib}

\providecommand{\latin}[1]{#1}
\makeatletter
\providecommand{\doi}
  {\begingroup\let\do\@makeother\dospecials
  \catcode`\{=1 \catcode`\}=2 \doi@aux}
\providecommand{\doi@aux}[1]{\endgroup\texttt{#1}}
\makeatother
\providecommand*\mcitethebibliography{\thebibliography}
\csname @ifundefined\endcsname{endmcitethebibliography}  {\let\endmcitethebibliography\endthebibliography}{}
\begin{mcitethebibliography}{28}
\providecommand*\natexlab[1]{#1}
\providecommand*\mciteSetBstSublistMode[1]{}
\providecommand*\mciteSetBstMaxWidthForm[2]{}
\providecommand*\mciteBstWouldAddEndPuncttrue
  {\def\EndOfBibitem{\unskip.}}
\providecommand*\mciteBstWouldAddEndPunctfalse
  {\let\EndOfBibitem\relax}
\providecommand*\mciteSetBstMidEndSepPunct[3]{}
\providecommand*\mciteSetBstSublistLabelBeginEnd[3]{}
\providecommand*\EndOfBibitem{}
\mciteSetBstSublistMode{f}
\mciteSetBstMaxWidthForm{subitem}{(\alph{mcitesubitemcount})}
\mciteSetBstSublistLabelBeginEnd
  {\mcitemaxwidthsubitemform\space}
  {\relax}
  {\relax}

\bibitem[van~den Hoven \latin{et~al.}(1993)van~den Hoven, Snoeks, Polman, van Uffelen, Oei, and Smit]{Al2O3_Si_Interface}
van~den Hoven,~G.~N.; Snoeks,~E.; Polman,~A.; van Uffelen,~J. W.~M.; Oei,~Y.~S.; Smit,~M.~K. Photoluminescence characterization of Er‐implanted Al2O3 films. \emph{Applied Physics Letters} \textbf{1993}, \emph{62}, 3065--3067\relax
\mciteBstWouldAddEndPuncttrue
\mciteSetBstMidEndSepPunct{\mcitedefaultmidpunct}
{\mcitedefaultendpunct}{\mcitedefaultseppunct}\relax
\EndOfBibitem
\bibitem[Smit \latin{et~al.}(1986)Smit, Acket, and {van der Laan}]{SMIT1986171}
Smit,~M.; Acket,~G.; {van der Laan},~C. Al2O3 films for integrated optics. \emph{Thin Solid Films} \textbf{1986}, \emph{138}, 171--181\relax
\mciteBstWouldAddEndPuncttrue
\mciteSetBstMidEndSepPunct{\mcitedefaultmidpunct}
{\mcitedefaultendpunct}{\mcitedefaultseppunct}\relax
\EndOfBibitem
\bibitem[Bonneville \latin{et~al.}(2024)Bonneville, Osornio-Martinez, Dijkstra, and Garc\'{i}a-Blanco]{Bonneville:24}
Bonneville,~D.~B.; Osornio-Martinez,~C.~E.; Dijkstra,~M.; Garc\'{i}a-Blanco,~S.~M. High on-chip gain spiral Al2O3:Er3\&\#x002B; waveguide amplifiers. \emph{Opt. Express} \textbf{2024}, \emph{32}, 15527--15536\relax
\mciteBstWouldAddEndPuncttrue
\mciteSetBstMidEndSepPunct{\mcitedefaultmidpunct}
{\mcitedefaultendpunct}{\mcitedefaultseppunct}\relax
\EndOfBibitem
\bibitem[Worhoff \latin{et~al.}(2009)Worhoff, Bradley, Ay, Geskus, Blauwendraat, and Pollnau]{4810183}
Worhoff,~K.; Bradley,~J. D.~B.; Ay,~F.; Geskus,~D.; Blauwendraat,~T.~P.; Pollnau,~M. Reliable Low-Cost Fabrication of Low-Loss $\hbox{Al}_{2}\hbox{O} _{3}{:}\hbox{Er}^{3+}$ Waveguides With 5.4-dB Optical Gain. \emph{IEEE Journal of Quantum Electronics} \textbf{2009}, \emph{45}, 454--461\relax
\mciteBstWouldAddEndPuncttrue
\mciteSetBstMidEndSepPunct{\mcitedefaultmidpunct}
{\mcitedefaultendpunct}{\mcitedefaultseppunct}\relax
\EndOfBibitem
\bibitem[Kenyon(2002)]{KENYON2002225}
Kenyon,~A. Recent developments in rare-earth doped materials for optoelectronics. \emph{Progress in Quantum Electronics} \textbf{2002}, \emph{26}, 225--284\relax
\mciteBstWouldAddEndPuncttrue
\mciteSetBstMidEndSepPunct{\mcitedefaultmidpunct}
{\mcitedefaultendpunct}{\mcitedefaultseppunct}\relax
\EndOfBibitem
\bibitem[Kik and Polman(1998)Kik, and Polman]{Kik_Polman_1998}
Kik,~P.; Polman,~A. Erbium-Doped Optical-Waveguide Amplifiers on Silicon. \emph{MRS Bulletin} \textbf{1998}, \emph{23}, 48–54\relax
\mciteBstWouldAddEndPuncttrue
\mciteSetBstMidEndSepPunct{\mcitedefaultmidpunct}
{\mcitedefaultendpunct}{\mcitedefaultseppunct}\relax
\EndOfBibitem
\bibitem[Ronn \latin{et~al.}(2016)Ronn, Karvonen, Kauppinen, Perros, Peyghambarian, Lipsanen, Saynatjoki, and Sun]{ronn2016atomic}
Ronn,~J.; Karvonen,~L.; Kauppinen,~C.; Perros,~A.~P.; Peyghambarian,~N.; Lipsanen,~H.; Saynatjoki,~A.; Sun,~Z. Atomic layer engineering of Er-ion distribution in highly doped Er: Al2O3 for photoluminescence enhancement. \emph{Acs Photonics} \textbf{2016}, \emph{3}, 2040--2048\relax
\mciteBstWouldAddEndPuncttrue
\mciteSetBstMidEndSepPunct{\mcitedefaultmidpunct}
{\mcitedefaultendpunct}{\mcitedefaultseppunct}\relax
\EndOfBibitem
\bibitem[Bradley \latin{et~al.}(2010)Bradley, Agazzi, Geskus, Ay, W\"{o}rhoff, and Pollnau]{Bradley:10}
Bradley,~J. D.~B.; Agazzi,~L.; Geskus,~D.; Ay,~F.; W\"{o}rhoff,~K.; Pollnau,~M. Gain bandwidth of 80 nm and 2 dB/cm peak gain in Al2O3:Er3$+$ optical amplifiers on silicon. \emph{J. Opt. Soc. Am. B} \textbf{2010}, \emph{27}, 187--196\relax
\mciteBstWouldAddEndPuncttrue
\mciteSetBstMidEndSepPunct{\mcitedefaultmidpunct}
{\mcitedefaultendpunct}{\mcitedefaultseppunct}\relax
\EndOfBibitem
\bibitem[Judd(1962)]{Judd}
Judd,~B.~R. Optical Absorption Intensities of Rare-Earth Ions. \emph{Phys. Rev.} \textbf{1962}, \emph{127}, 750--761\relax
\mciteBstWouldAddEndPuncttrue
\mciteSetBstMidEndSepPunct{\mcitedefaultmidpunct}
{\mcitedefaultendpunct}{\mcitedefaultseppunct}\relax
\EndOfBibitem
\bibitem[Ofelt(1962)]{Ofelt}
Ofelt,~G.~S. Intensities of Crystal Spectra of Rare‐Earth Ions. \emph{The Journal of Chemical Physics} \textbf{1962}, \emph{37}, 511--520\relax
\mciteBstWouldAddEndPuncttrue
\mciteSetBstMidEndSepPunct{\mcitedefaultmidpunct}
{\mcitedefaultendpunct}{\mcitedefaultseppunct}\relax
\EndOfBibitem
\bibitem[Khan and Leuenberger(2021)Khan, and Leuenberger]{ER_W_khan}
Khan,~M.~A.; Leuenberger,~M.~N. Ab initio calculations for electronic and optical properties of Er$_W$ defects in single-layer tungsten disulfide. \emph{Journal of Applied Physics} \textbf{2021}, \emph{130}, 115104\relax
\mciteBstWouldAddEndPuncttrue
\mciteSetBstMidEndSepPunct{\mcitedefaultmidpunct}
{\mcitedefaultendpunct}{\mcitedefaultseppunct}\relax
\EndOfBibitem
\bibitem[Khan and Leuenberger(2022)Khan, and Leuenberger]{khan2022first}
Khan,~M.; Leuenberger,~M.~N. First-principles study of the electronic and optical properties of Ho$_W$ impurities in single-layer tungsten disulfide. \emph{Scientific Reports} \textbf{2022}, \emph{12}, 11437\relax
\mciteBstWouldAddEndPuncttrue
\mciteSetBstMidEndSepPunct{\mcitedefaultmidpunct}
{\mcitedefaultendpunct}{\mcitedefaultseppunct}\relax
\EndOfBibitem
\bibitem[Khan \latin{et~al.}(2017)Khan, Erementchouk, Hendrickson, and Leuenberger]{Khan2017}
Khan,~M.~A.; Erementchouk,~M.; Hendrickson,~J.; Leuenberger,~M.~N. Electronic and optical properties of vacancy defects in single-layer transition metal dichalcogenides. \emph{Phys. Rev. B} \textbf{2017}, \emph{95}, 245435\relax
\mciteBstWouldAddEndPuncttrue
\mciteSetBstMidEndSepPunct{\mcitedefaultmidpunct}
{\mcitedefaultendpunct}{\mcitedefaultseppunct}\relax
\EndOfBibitem
\bibitem[Erementchouk \latin{et~al.}(2015)Erementchouk, Khan, and Leuenberger]{Erementchouk2015}
Erementchouk,~M.; Khan,~M.~A.; Leuenberger,~M.~N. Optical signatures of states bound to vacancy defects in monolayer ${\mathrm{MoS}}_{2}$. \emph{Phys. Rev. B} \textbf{2015}, \emph{92}, 121401\relax
\mciteBstWouldAddEndPuncttrue
\mciteSetBstMidEndSepPunct{\mcitedefaultmidpunct}
{\mcitedefaultendpunct}{\mcitedefaultseppunct}\relax
\EndOfBibitem
\bibitem[Snitzer and Woodcock(1965)Snitzer, and Woodcock]{Er_Glass_Laser}
Snitzer,~E.; Woodcock,~R. Yb$^{3+}$–Er$^3+$ GLASS LASER. \emph{Applied Physics Letters} \textbf{1965}, \emph{6}, 45--46\relax
\mciteBstWouldAddEndPuncttrue
\mciteSetBstMidEndSepPunct{\mcitedefaultmidpunct}
{\mcitedefaultendpunct}{\mcitedefaultseppunct}\relax
\EndOfBibitem
\bibitem[Malinowski \latin{et~al.}(2000)Malinowski, Frukacz, Szuflińska, Wnuk, and Kaczkan]{Malinowski2000}
Malinowski,~M.; Frukacz,~Z.; Szuflińska,~M.; Wnuk,~A.; Kaczkan,~M. Optical transitions of Ho$^{3+}$ in YAG. \emph{Journal of Alloys and Compounds} \textbf{2000}, \emph{300-301}, 389--394\relax
\mciteBstWouldAddEndPuncttrue
\mciteSetBstMidEndSepPunct{\mcitedefaultmidpunct}
{\mcitedefaultendpunct}{\mcitedefaultseppunct}\relax
\EndOfBibitem
\bibitem[Aisaka \latin{et~al.}(2008)Aisaka, Fujii, and Hayashi]{aisaka2008enhancement}
Aisaka,~T.; Fujii,~M.; Hayashi,~S. Enhancement of upconversion luminescence of Er doped Al2O3 films by Ag island films. \emph{Applied Physics Letters} \textbf{2008}, \emph{92}\relax
\mciteBstWouldAddEndPuncttrue
\mciteSetBstMidEndSepPunct{\mcitedefaultmidpunct}
{\mcitedefaultendpunct}{\mcitedefaultseppunct}\relax
\EndOfBibitem
\bibitem[Xiao and Wang(2022)Xiao, and Wang]{xiao2022design}
Xiao,~P.; Wang,~B. Design of an erbium-doped Al2O3 optical waveguide amplifier with on-chip integrated laser pumping source. \emph{Optics Communications} \textbf{2022}, \emph{508}, 127709\relax
\mciteBstWouldAddEndPuncttrue
\mciteSetBstMidEndSepPunct{\mcitedefaultmidpunct}
{\mcitedefaultendpunct}{\mcitedefaultseppunct}\relax
\EndOfBibitem
\bibitem[Perdew \latin{et~al.}(1996)Perdew, Burke, and Ernzerhof]{PhysRevLett.77.3865}
Perdew,~J.~P.; Burke,~K.; Ernzerhof,~M. Generalized Gradient Approximation Made Simple. \emph{Phys. Rev. Lett.} \textbf{1996}, \emph{77}, 3865--3868\relax
\mciteBstWouldAddEndPuncttrue
\mciteSetBstMidEndSepPunct{\mcitedefaultmidpunct}
{\mcitedefaultendpunct}{\mcitedefaultseppunct}\relax
\EndOfBibitem
\bibitem[QW_(2021.6)]{QW_1}
VNL-ATK 2019.12. \emph{http://www.quantumwise.com/} \textbf{2021.6}, \relax
\mciteBstWouldAddEndPunctfalse
\mciteSetBstMidEndSepPunct{\mcitedefaultmidpunct}
{}{\mcitedefaultseppunct}\relax
\EndOfBibitem
\bibitem[Tran and Blaha(2009)Tran, and Blaha]{meta_GGA}
Tran,~F.; Blaha,~P. Accurate Band Gaps of Semiconductors and Insulators with a Semilocal Exchange-Correlation Potential. \emph{Phys. Rev. Lett.} \textbf{2009}, \emph{102}, 226401\relax
\mciteBstWouldAddEndPuncttrue
\mciteSetBstMidEndSepPunct{\mcitedefaultmidpunct}
{\mcitedefaultendpunct}{\mcitedefaultseppunct}\relax
\EndOfBibitem
\bibitem[Galasso(2013)]{galasso2013structure}
Galasso,~F.~S. \emph{Structure and properties of inorganic solids: international series of monographs in solid state physics}; Elsevier, 2013; Vol.~7\relax
\mciteBstWouldAddEndPuncttrue
\mciteSetBstMidEndSepPunct{\mcitedefaultmidpunct}
{\mcitedefaultendpunct}{\mcitedefaultseppunct}\relax
\EndOfBibitem
\bibitem[French(1990)]{Exp_BG_1}
French,~R.~H. Electronic Band Structure of Al2O3, with Comparison to Alon and AIN. \emph{Journal of the American Ceramic Society} \textbf{1990}, \emph{73}, 477--489\relax
\mciteBstWouldAddEndPuncttrue
\mciteSetBstMidEndSepPunct{\mcitedefaultmidpunct}
{\mcitedefaultendpunct}{\mcitedefaultseppunct}\relax
\EndOfBibitem
\bibitem[Bortz and French(1989)Bortz, and French]{Exp_BG_2}
Bortz,~M.~L.; French,~R.~H. Optical reflectivity measurements using a laser plasma light source. \emph{Applied Physics Letters} \textbf{1989}, \emph{55}, 1955--1957\relax
\mciteBstWouldAddEndPuncttrue
\mciteSetBstMidEndSepPunct{\mcitedefaultmidpunct}
{\mcitedefaultendpunct}{\mcitedefaultseppunct}\relax
\EndOfBibitem
\bibitem[Khan \latin{et~al.}(2025)Khan, Atif, and Leuenberger]{Impurity_Enginneered_Graphullerene}
Khan,~M.~A.; Atif,~M.; Leuenberger,~M.~N. Tuning the electronic and optical properties of impurity-engineered two-dimensional graphullerene half-semiconductors. \emph{Phys. Rev. Mater.} \textbf{2025}, \emph{9}, 034001\relax
\mciteBstWouldAddEndPuncttrue
\mciteSetBstMidEndSepPunct{\mcitedefaultmidpunct}
{\mcitedefaultendpunct}{\mcitedefaultseppunct}\relax
\EndOfBibitem
\bibitem[V\'{a}zquez-C\'{o}rdova \latin{et~al.}(2014)V\'{a}zquez-C\'{o}rdova, Dijkstra, Bernhardi, Ay, W\"{o}rhoff, Herek, Garc\'{i}a-Blanco, and Pollnau]{Vazquez-Cordova:14}
V\'{a}zquez-C\'{o}rdova,~S.~A.; Dijkstra,~M.; Bernhardi,~E.~H.; Ay,~F.; W\"{o}rhoff,~K.; Herek,~J.~L.; Garc\'{i}a-Blanco,~S.~M.; Pollnau,~M. Erbium-doped spiral amplifiers with 20 dB of net gain on silicon. \emph{Opt. Express} \textbf{2014}, \emph{22}, 25993--26004\relax
\mciteBstWouldAddEndPuncttrue
\mciteSetBstMidEndSepPunct{\mcitedefaultmidpunct}
{\mcitedefaultendpunct}{\mcitedefaultseppunct}\relax
\EndOfBibitem
\bibitem[Yang \latin{et~al.}(2007)Yang, Dai, and Sun]{YANG2007207}
Yang,~H.; Dai,~Z.; Sun,~Z. Upconversion luminescence and kinetics in Er3+:YAlO3 under 652.2nm excitation. \emph{Journal of Luminescence} \textbf{2007}, \emph{124}, 207--212\relax
\mciteBstWouldAddEndPuncttrue
\mciteSetBstMidEndSepPunct{\mcitedefaultmidpunct}
{\mcitedefaultendpunct}{\mcitedefaultseppunct}\relax
\EndOfBibitem
\end{mcitethebibliography}
\end{document}